\documentclass[12pt]{iopart}

\usepackage{graphicx}% Include figure files
\usepackage{dcolumn}% Align table columns on decimal point
\usepackage{bm}%
\usepackage{amsmath}
\usepackage{caption}
\usepackage[T1]{fontenc}

\usepackage{hyperref}
\hypersetup{
    colorlinks=true,
    linkcolor=blue,
    filecolor=blue,      
    urlcolor=blue,
    citecolor=blue
}

\begin{document}

\title{Interface and electromagnetic effects in the valley splitting of Si quantum dots}

\author{Jonas R F Lima$^{1,2}$ and Guido Burkard$^1$}
\address{$^1$Department of Physics, University of Konstanz, 78457 Konstanz, Germany}
\address{$^2$Departamento de F\'{\i}sica, Universidade Federal Rural de Pernambuco, 52171-900, Recife, PE, Brazil}
\ead{jonas.de-lima@uni-konstanz.de, guido.burkard@uni-konstanz.de}

\vspace{10pt}
\begin{indented}
\item[]February 2023
\end{indented}

\begin{abstract}

The performance and scalability of silicon spin qubits depend directly on the value of the conduction band valley splitting. In this work, we investigate the influence of electromagnetic fields and the interface width on the valley splitting of a quantum dot in a Si/SiGe heterostructure. We propose a new three-dimensional theoretical model within the effective mass theory for the calculation of the valley splitting in such heterostructures that takes into account the concentration fluctuation at the interfaces and the lateral confinement. With this model, we predict that the electric field is an important parameter for valley splitting engineering, since it can shift the probability distribution away from small valley splittings for some interface widths. We also obtain a critical softness of the interfaces in the heterostructure, above which the best option for spin qubits is to consider an interface as wide as possible. 

\end{abstract}

\section{Introduction}

The weak spin-orbit coupling and the nuclear zero-spin isotopes of silicon and germanium make Si/Ge quantum dots (QD) an ideal host for semiconductor spin qubits \cite{RevModPhys.85.961,Burkard2021arXiv}, allowing for long relaxation \cite{Morello,Yang,PhysRevApplied.11.044063} and dephasing \cite{PhysRevB.83.165301,Tyryshkin,doi:10.1126/science.1217635} times. Experiments in silicon structures have demonstrated high fidelities for single and two-qubit gates \cite{Veldhorst,Yoneda,doi:10.1126/science.aao5965,Watson,Huang,Xue,Noiri}, entanglement of three-spin states \cite{Takeda2021}, and a strong coherent coupling between a single spin and single microwave-frequency photons \cite{Mi2018,doi:10.1126/science.aar4054}. However, the degeneracy of the conduction band minima (valleys) of bulk silicon limits the performance of quantum information processing because the less coherent valley degree of freedom competes with the spin as a low-energy two-level system. The valley degeneracy is lifted in quantum dots in Si/SiGe heterostructures due to biaxial strain and a sharp interface potential, but the reported valley splittings are often uncontrolled and can be as low as 10 to 100 $\mu$eV, close to the thermal energy $k_B T$ as cryogenic temperatures $T\approx 0.1-1$K and the spin (Zeeman) splitting $g\mu_B B$ at $B\approx 1$T \cite{doi:10.1063/1.3569717,doi:10.1063/1.3666232,doi:10.1063/1.4922249,PhysRevApplied.13.034068,PhysRevApplied.15.044033,PhysRevB.95.165429,PhysRevB.98.161404,PhysRevLett.119.176803}. One manifestation of the valley degeneracy consists in a very fast spin relaxation which is observed when the valley splitting becomes equal to the qubit Zeeman splitting, a phenomenon known as spin-valley hotspot \cite{Yang,PhysRevB.90.235315}. On top of that, the valley splitting (VS) of devices fabricated on the same heterostructure growth vary wildly \cite{PaqueletWuetz2022}. Such variability is a consequence of the random concentration fluctuations at the Si/SiGe interfaces, which poses a challenge for the control of the valley splitting.

The greatest values for the valley splitting are obtained for very sharp interfaces with a thinkness of $\leq$ 2 monolayers (ML), but the fabrication of such interfaces is not realistic. Therefore, different proposals for the enhancement of the valley splitting in Si-based heterostructures were reported recently. Some of these proposals are related to the increase of the Ge content inside of the Si quantum well, which can be done, e.g., by introducing a low concentration of Ge \cite{PaqueletWuetz2022}, a single Ge layer \cite{PhysRevB.104.085406}, or an oscillating Ge concentration \cite{McJunkin2022,PhysRevB.106.085304}. The enhancement of the valley splitting in these proposals is related to the increase of the overlap between the electron wavefunction and the Ge atoms. In Ref.~\cite{PaqueletWuetz2022} it was also suggested that a nonintuitive increase of the interface width leads to an enhancement of the valley splitting. However, a complete analysis of the valley splitting as a function of the interface width is missing. As a consequence, it is an open question whether a realistically sharp interface or a very wide interface is better for spin qubit applications. 

The valley splitting in Si-based heterostructures can be calculated either using atomistic techniques \cite{doi:10.1063/1.1637718,4294217,doi:10.1063/1.2591432} or effective mass theory \cite{PhysRevB.75.115318,PhysRevB.84.155320,PhysRevB.86.035321,PhysRevB.100.125309,PhysRevB.80.081305,doi:10.1063/1.2387975}. While atomistic methods can in principle work with very few free parameters, the advantage of effective-mass theories is that the results are straightforward and, in some cases, analytical, while still agreeing with the other methods. The valley splitting can be obtained from the envelope function of the electronic wavefunction, and the effective mass theory is used to deduce this envelope function. Several works have considered variational methods or an infinite potential barrier to obtain the envelope function, but these methods are not accurate, e.g., when a strong electric field is present. The influence of electromagnetic fields in the valley splitting was predicted in Ref.~\cite{PhysRevResearch.2.043180}, where a Si/SiGe interface and also an interface between SiGe and an insulator layer hosting the gate electrodes were considered. However, the calculations were restricted to an ideal step interface. A one-dimensional model that takes into account the concentration fluctuations at the Si/SiGe interfaces was proposed recently \cite{PaqueletWuetz2022}, which does not take into consideration the influence of the shape and location of the quantum dot.

In this work we propose a new three-dimensional model within the effective mass theory for the calculation of the valley splitting of quantum dots in silicon-based heterostructures. The main idea of the model is to build up the confinement potential along the quantum dot growth direction by delta functions at the location of each Ge atom. In this way, we can model the concentration fluctuations at the Si/SiGe interfaces and also take into account the lateral confinement in the calculation of the valley splitting, since we have now a three dimensional potential profile given by the delta functions that are distributed in the three directions. With this model, we calculate the valley splitting as a function of the interface width and electromagnetic fields. Analysing the statistics of the valley splitting, we show that the electric field is an important parameter for valley splitting engineering since it can, e.g., shift up the distribution of valley splitting for some interface widths. We also obtain a critical interface width, in such way that the best configuration for spin qubit applications is an interface narrower than this critical value. However, if such sharpness cannot be achieved, the best option is to fabricate devices with an interface that is as wide as possible. Additionally, we obtain that an in-plane magnetic field has a very weak influence in the valley splitting.

The remainder of this paper is organised as follows. In Sec.~\ref{sec:model}, we explain in detail the theoretical model that we are proposing here and use it to obtain the envelope function of states confined to the quantum dot in the SiGe/Si/SiGe heterosctructure. We consider first the case without a magnetic field, Sec.~\ref{subsec:nofield}, and then with the presence of an in-plane magnetic field, Sec.~\ref{subsec:field}. In Sec.~\ref{sec:results}, we obtain our results and discuss the influence of the interface width and electromagnetic fields in the valley splitting. The paper is summarised and concluded in Sec.~\ref{sec:conclusion}. Two appendices were included to add more technical details about our theoretical model and the numerical calculations performed here.  

\section{Model}
\label{sec:model}

\begin{figure}[h]
\center
\includegraphics[width=0.8\linewidth]{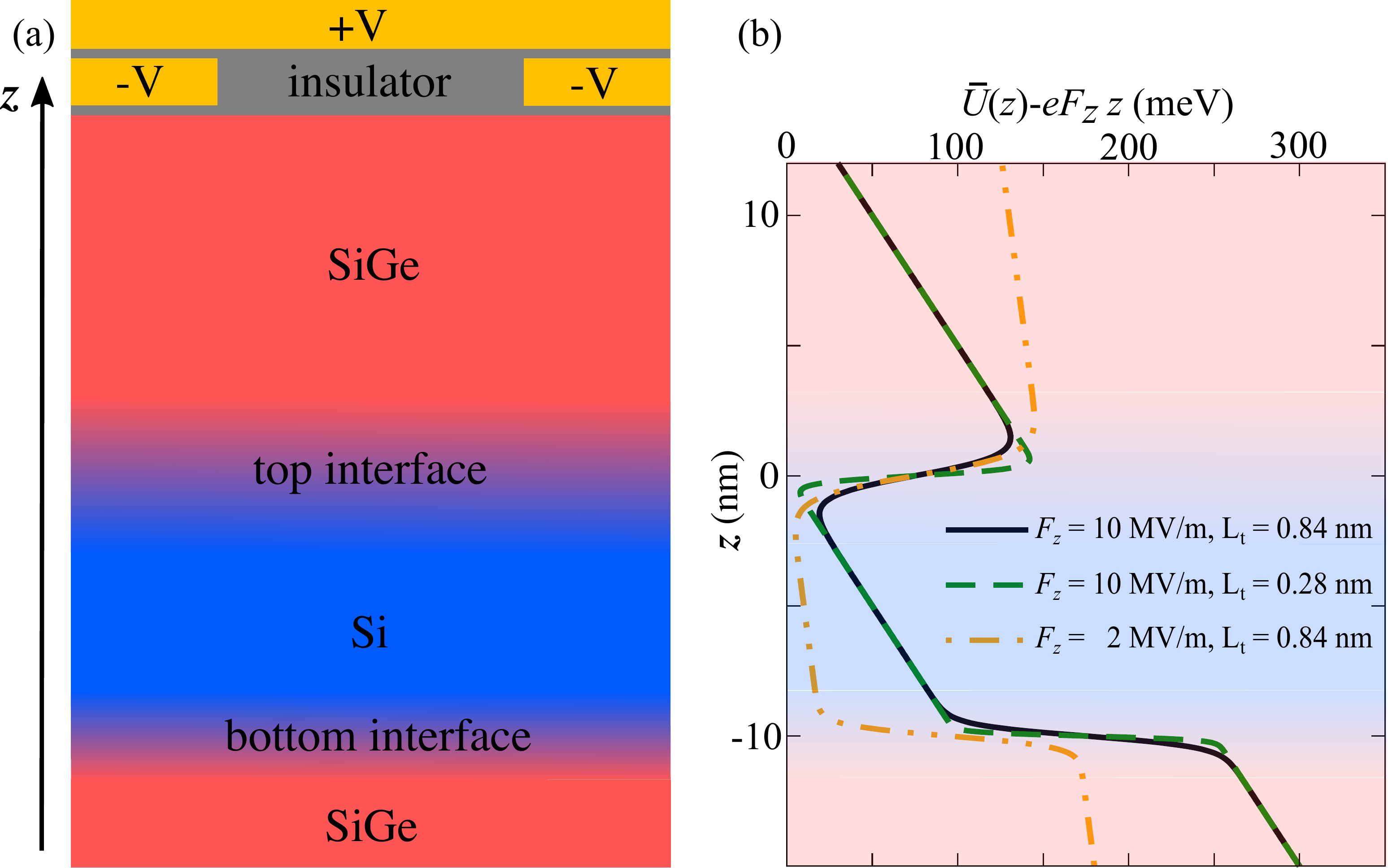}
\caption{\small (a) Schematic cross-section of the SiGe/Si/SiGe heterostructure containing the quantum dot. The system is divided into six regions: the lower and upper SiGe barriers, the Si well, the top and bottom interfaces and the insulating layer. The $\pm V$ gates are responsible for the lateral confinement of the quantum dot. (b) Th average electrostatic potential along the $\hat{z}$ direction for two distinct electric fields and interface widths, where $F_z$ denotes the electric field and $\bar{U}$ the average potential induced by the potential barriers. In all cases we consider $L_b = L_t/2$.} \label{system}
\end{figure}

We consider here a SiGe/Si/SiGe heterostructure that is grown along the $\hat{z}$ direction, as can be seen in the schematic cross-section in Fig.  \ref{system} $(a)$. A realistic heterostructure does not comprise ideal step Si/SiGe interfaces, which were considered in previous works \cite{PhysRevB.75.115318,PhysRevResearch.2.043180}. Rather than an abrupt change in Ge concentration, there is a smooth transition in the Ge concentration along the z direction. Having this in mind, we divide the system in six regions: the upper and lower SiGe barriers, the Si well, the top and bottom interfaces between Si and SiGe and the insulating region. Also, gates are used to trap and confine electrons in the silicon layer and to induce an electric field in the $\hat{z}$ direction. We consider, e.g., a Si layer of $d_w = 10$~nm that is located at $-d_w \leq z \leq 0$ and that the interface between the upper SiGe barrier and the insulator region is at $z = d_i = 46$~nm.

The electron wavefunction can probe the random distribution of the Si and Ge atoms in the SiGe alloy within the Si/SiGe interface region. Therefore, sample-to-sample fluctuations in the Ge concentration are relevant for the electronic properties, and in particular for the distribution of valley splittings. We propose here a new model for the calculation of the valley splitting in such devices. The main idea is to replace, in the top and bottom interfaces, the confinement potential in the $z$ direction by a delta function potential at the location of each Ge atom. Since the Ge atoms are distributed in three dimensions, the confinement potential $U(x,y,z)$ depends on all three spatial coordinates $x$, $y$, and $z$. The Hamiltonian in the absence of a magnetic field that describes the envelope function of the system is given by
\begin{eqnarray}
H=\frac{p_x^2}{2m_t} + \frac{1}{2} m_t \omega_x^2 x^2 + \frac{p_y^2}{2m_t} + \frac{1}{2} m_t \omega_y^2 y^2 + \frac{p_z^2}{2m_l} -eF_zz + U(x,y,z),
\label{H}
\end{eqnarray}
where $m_t = 0.19 \; m_e$ and $m_l = 0.98 \; m_e$ are the transverse and longitudinal effective masses and $\omega_x = 2\hbar/m_tx_0^2$ and $\omega_y = 2\hbar/m_ty_0^2$ are the confinement frequencies along $\hat{x}$ and $\hat{y}$ directions, with $x_0$ and $y_0$ being the size (radius or semi-axis) of the quantum dot along $\hat{x}$ and $\hat{y}$.

We can separate the potential $U$ in the six regions of the system.  It will depend on $x$ and $y$ only in the top and bottom interfaces. We assume that the SiGe regions have 30$\%$ of germanium on average, which is a typical value. In this case, the energy offset between the conduction band minima in Si and SiGe is 150 meV.  So, the potential in the SiGe barriers is $U_0 = 150$~meV, while it is 0 at the silicon well region. In the insulator region we take $U \rightarrow \infty$. Within the interfaces, we have that
\begin{equation}
U_I(x,y,z)=\lambda \sum_i \delta(x-x_i) \delta(y-y_i) \delta(z-z_i),
\label{Ui}
\end{equation}
where $i$ labels the Ge atoms in the interface region and $\lambda$ is a parameter of the model that quantifies the strength of the interaction between a Ge impurity and a conduction-band electron.

We consider in all the results obtained here a total of $10^4$ realizations of the random Ge atom positions $(x_i,y_i,z_i)$. Each realization represents a random distribution of the Ge atoms at the interface. We assume that in the $x$ and $y$ directions the Ge atoms are uniformly distributed, while in the $z$ direction they are distributed following a probability distribution function (PDF) given by a hyperbolic tangent function (further details in \ref{sec:potential}). This means that we can replace the potential (\ref{Ui}) by an average potential $\bar{U}$, which is taken over the $10^4$ realizations, plus a fluctuation $\delta U$. Since the Ge atoms are distributed uniformly in the $x$ and $y$ direction, $\bar{U}$ is constant in these directions. So, we have now that at the interface,
\begin{eqnarray}
U_I(x,y,z)= \bar{U}(z)+\delta U(x,y,z).
\end{eqnarray}
where
\begin{eqnarray}
\bar{U}(z)= \frac{U_0}{2}[\tanh ((-d_w-z)/L_b)+1]+\frac{U_0}{2}[\tanh (z/L_t)+1],
\label{ubar}
\end{eqnarray}
where $L_b$ and $L_t$ control the widths of the bottom and top interfaces, respectively. In Fig.~\ref{system} $(b)$ we plot $\bar{U}$ for two values of $L_t$ and $F_z$, with $L_b = L_t/2$ in all cases. Even though we are modelling the smooth interface by a hyperbolic tangent function, other functions can also be used \cite{PhysRevB.80.081305}.

It is important to mention that $\lambda$ is not a free parameter of the model. In fact, it is fixed in such a way that the average potential from various realizations of the potential (\ref{Ui}) reproduces the potential (\ref{ubar}). We find that $\lambda = 10$~meV$\cdot$nm reproduces the conduction band offset of $U_0 = 150$~meV at the Si/Si$_{0.7}$Ge$_{0.3}$ interface. More details can be seen in the \ref{sec:potential}. Furthermore, we consider in all results an elliptical quantum dot with $x_0 = 12$~nm and $y_0 = 15$~nm.

\subsection{Envelope function without magnetic field}
\label{subsec:nofield}

In order to calculate the valley splitting using the effective mass theory, we first need to obtain the envelope function. Considering $\delta U(x,y,z)$ as a perturbation, we can obtain the (unperturbed) envelope function from the Hamiltonian (\ref{H}) using separation of variables, in such a way that $\psi_{xyz} = \psi_x\psi_y\psi_z$, and $\psi_x(x)$ and $\psi_y(y)$ are harmonic oscillator wavefunctions, which have well-known eigenenergies $E_{x,n_x}$ and $E_{y,n_y}$. We then need to obtain the eigenstates $\psi_z(z)$ and eingenenergies $E_{z,n_z}$ for the electron motion in the $z$ direction.

The Schr\"odinger equation for the envelope function in the $z$ direction is given by
\begin{eqnarray}
\left( \frac{p_z^2}{2m_l} -eF_zz + \bar{U}(z) - E_{z,n_z} \right)\psi_{z,n_z}=0.
\label{sch}
\end{eqnarray}
Using the electrical confinement length
\begin{eqnarray}
z_0 = \left(\frac{\hbar^2}{2m_leF_z}\right)^{1/3},
\end{eqnarray}
and the energy scale
\begin{eqnarray}
\epsilon_0 = \frac{\hbar^2}{2m_lz_0^2},
\end{eqnarray}
we can rewrite Eq. (\ref{sch}) as
\begin{eqnarray}
\left[ \frac{d^2}{d\tilde{z}^2} -(\tilde{U} - \tilde{z}   - \tilde{\epsilon}_{z,n_z}) \right]\psi_{z,n_z}=0.
\label{sch2}
\end{eqnarray}
where $\tilde{U} = \bar{U}/\epsilon_0$, $\tilde{z} = z/z_0$ and $\tilde{\epsilon}_{z,n_z} = \epsilon_{z,n_z}/\epsilon_0$.

For a constant $\tilde{U}$, the analytical solution of the above equation is a linear combination of the Airy functions of first and second kind. We solve this equation numerically using the transfer matrix method, where we decompose the potential in successive constant rectangular barriers and use the continuity of the wave function and its first derivative at each interface (further details in \ref{sec:numeric}).   

\begin{figure}[h]
\center
\includegraphics[width=0.8\linewidth]{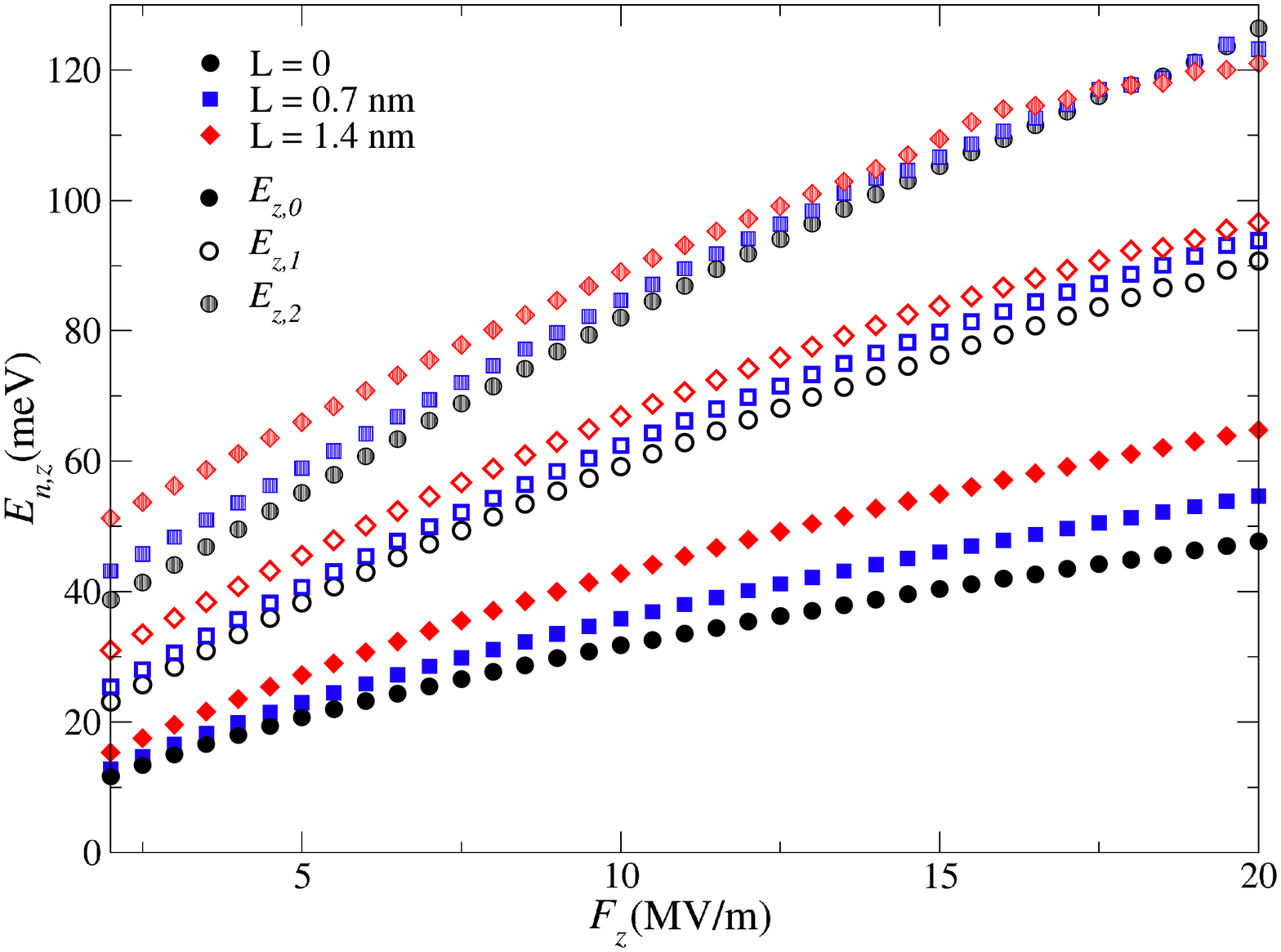}
\caption{\small The eigenenergies of the ground state $E_{z,0}$ and the first two excited states, $E_{z,1}$ and $E_{z,2}$, as a function of the electric field $F_z$ for three different values of $L$ (interface width is equal to $5.4\cdot L$).} \label{energy}
\end{figure}

In Fig. \ref{energy} we show the eigenenergies $E_{z,n_z}$ for the ground state $n_z=0$ and the first two excited states $n_z=1,2$ as a function of the electric field $F_z$ for three different values of $L$, where we are considering $L_b = L_t = L$. The case with $L = 0$ means an ideally sharp step interface. The geometric width of the interface is given by $5.4\cdot L$ (see \ref{sec:potential}). Besides these energies, we also have a set of states whose envelope function is localized in the upper SiGe barrier. This is a consequence of the quantum well created by the electric field between the top interface and the insulator region. We neglect these states because they do not contribute to the behaviour of the valley splitting \cite{PhysRevResearch.2.043180}.  

The increase in the energies as we increase the width of the interface can be explained by the fact that the bottom of the quantum well at the silicon layer is shifted up for a smooth interface. This can be seen clearly around $z=0$ in Fig.~\ref{system}(b). Also, the height of the potential barrier at the top interface is shifted down when we increase the interface width and/or the electric field. For this reason, the second excited state for a strong electric field and smooth interfaces is no longer confined to the silicon quantum well, which explaing the lower energy for these states compared to the step case ($L = 0$).

\begin{figure}[h]
\center
\includegraphics[width=0.8\linewidth]{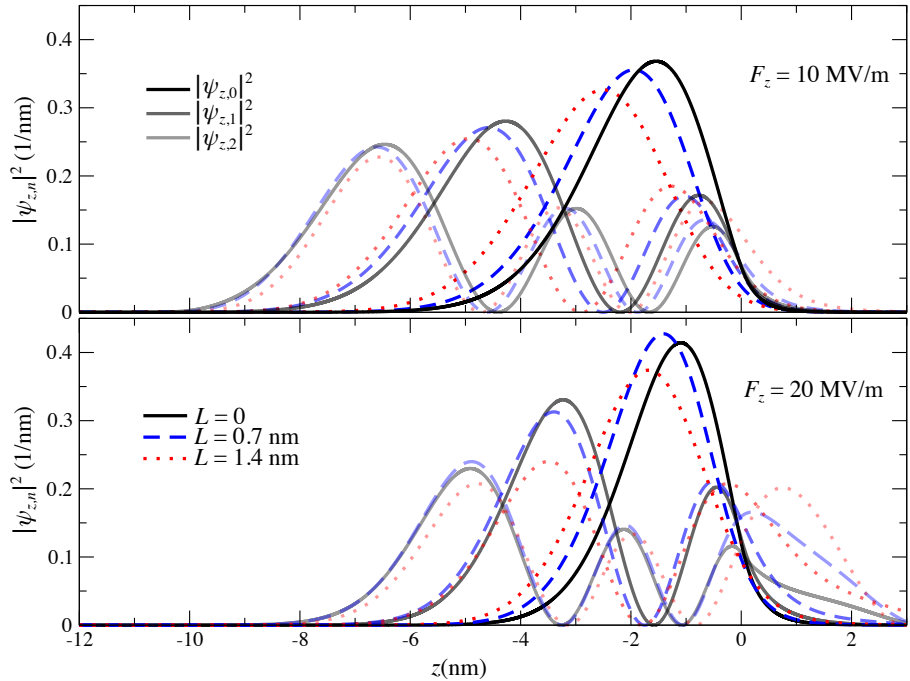}
\caption{\small The probability density along $\hat{z}$ direction $|\psi_{z,n}|^2$ of the ground state and the first two excited states for three different values of $L$. We consider in the top and bottom panels an electric field of 10 MV/M and 20 MV/m, respectively.} \label{envelope}
\end{figure}

We also show the probability density $|\psi_{z,n}|^2$ in Fig. \ref{envelope} for the ground state and the first two excited states. We can see that the electric field pushes the envelope function towards the upper SiGe barrier. This is important for the calculation of the valley splitting, where we will neglect the probability density at the bottom interface and lower SiGe barrier for strong electric fields. Also, we note only a small penetration of the envelope function in the upper SiGe barrier, which is not the case when the eigenstate is not confined in the Si well, as can be seen for $F_z = 20$~MV/m and $L = 1.4$~nm. 

It is important to mention that the envelope functions were obtained from the averaged potential $\bar{U}_z$, while the corrections due to the fluctuations $\delta U$ are neglected in this step. However, $\delta U$ will play an important role when evaluating the valley splitting.

\subsection{Envelope function in the presence of a magnetic field}
\label{subsec:field}

We consider now an in-plane magnetic field $\boldsymbol{B} = (B_x,B_y,0)$. We calculate the envelope function here as it was done in Ref. \cite{PhysRevResearch.2.043180}. With the vector potential $\boldsymbol{A}= (0,0,yB_x-xB_y)$, Eq. (\ref{H}) can be written as
\begin{eqnarray}
    H =  H^{0}(\boldsymbol{B}) + H^{\prime}(\boldsymbol{B}),
\end{eqnarray}
where $H^{0}(\boldsymbol{B})$ is equal to the Hamiltonian in Eq. (\ref{H}) with the substitution
\begin{eqnarray}
    \omega_{x(y)} \rightarrow \omega_{x(y)}\left( 1+\frac{\Omega^2_{y(x)}}{\omega^2_{x(y)}} \right)^{1/2} = \omega_{x(y)}^{\prime},
\end{eqnarray}
with the cyclotron frequency given by
\begin{eqnarray}
    \Omega_{x(y)} = \frac{eB_{x(y)}}{\sqrt{m_lm_t}}.
\end{eqnarray}
We treat 
\begin{eqnarray}
    H^{\prime}(\boldsymbol{B})= -\frac{e}{m_l}(yB_x-xB_y)p_z - \frac{e^2}{m_l}B_xB_yxy
\end{eqnarray}
as a perturbation and consider only first order correction to the envelope function.

We solve $H^{0}(\boldsymbol{B})$ as it was done in the previous section to obtain the unperturbed envelope function $\psi_{xyz}^{0}$. The only difference is that the harmonic oscillator solutions at $\hat{x}$ and $\hat{y}$ directions have now the magnetic field dependent confinement frequencies $\omega_{x(y)}^{\prime}$. The solution at $\hat{z}$ direction is not affected by $\boldsymbol{B}$. The correction to the envelope function due to $H^{\prime}(\boldsymbol{B})$ is given by
\begin{eqnarray}
    \psi_{xyz}^{\prime}=-iB_x\frac{y_0^{\prime}}{z_0}\psi_{x,0}\psi_{y,1}\sum_{\substack{m=0\\m\neq n_z}}^{m_{\text{max}}}\alpha^{y}_m \psi_{z,m} &+& iB_y\frac{x_0^{\prime}}{z_0}\psi_{x,1}\psi_{y,0}\sum_{\substack{m=0\\m\neq n_z}}^{m_{\text{max}}}\alpha^{x}_m \psi_{z,m} \nonumber \\
    &+&\frac{B_xB_yx_0^{\prime}y^{\prime}_0e^2}{4m_l\hbar(\omega^{\prime}_x+\omega^{\prime}_y)}\psi_{x,1}\psi_{y,1}\psi_{z,n_z},
\end{eqnarray}
where the sum is over all states confined to the Si quantum well, which means that $m_{\text{max}}$ depends on $F_z$ and $L$, and 
\begin{eqnarray}
    x^{\prime}_0 = x_0\left(1+\frac{m_tx_0^4}{4m_ll_y^4}\right)^{-1/4},\quad
    y^{\prime}_0 =  y_0\left(1+\frac{m_ty_0^4}{4m_ll_x^4}\right)^{-1/4},
\end{eqnarray}
with the magnetic length given by
\begin{eqnarray}
    l_{x(y)}=\sqrt{\frac{\hbar}{eB_{x(y)}}}.
\end{eqnarray}
For the coefficients $\alpha^{x(y)}$,
we find
\begin{eqnarray}
    \alpha^{x(y)}_m = \frac{e\hbar}{2m_l} \frac{\langle \psi_{z,n_z}|\partial/\partial \tilde{z}|\psi_{z,m}\rangle}{E_{z,n_z}-E_{z,m}-\omega^{\prime}_{x(y)}}.
\end{eqnarray}

\section{Results}
\label{sec:results}

The two low-lying valley states of the quantum dot are given by
\begin{equation}
|\pm z\rangle = \Psi_{xyz}e^{\pm ik_0z}u_{\pm z}(\textbf{r}),
\end{equation}
where $ \Psi_{xyz}$ is the envelope function, $u_{\pm z}(\textbf{r})$ are the periodic parts of the Bloch functions, $k_0 = 0.82(2\pi /a_0)$ is the Bloch wavenumber at the conduction band minima of silicon and $a_0 = 0.543$~nm is the length of the Si cubic unit cell. 

The intervalley coupling is given by
\begin{equation}
\Delta = \langle +z | -eF_zz+U(x,y,z)|-z\rangle.
\end{equation} 
As already obtained in previous works, the contribution to the intervalley coupling from the electric field is negligibly small compared to the other term. So, we can write
\begin{eqnarray}
\Delta=C_0\int e^{-2ik_0z}U(x,y,z)|\psi_{x,y,z}|^2 d^3x,
\end{eqnarray}
where $C_0 = -0.2607$ comes from the periodic parts of the Bloch wavefunctions \cite{PhysRevB.84.155320,PhysRevResearch.2.043180}. 
The total valley splitting is then given by
\begin{eqnarray}
E_{\rm VS} = 2 |\Delta|.
\end{eqnarray}

\subsection{Valley splitting in the absence of a magnetic field}

\begin{figure}[]
\center
\includegraphics[width=0.8\columnwidth]{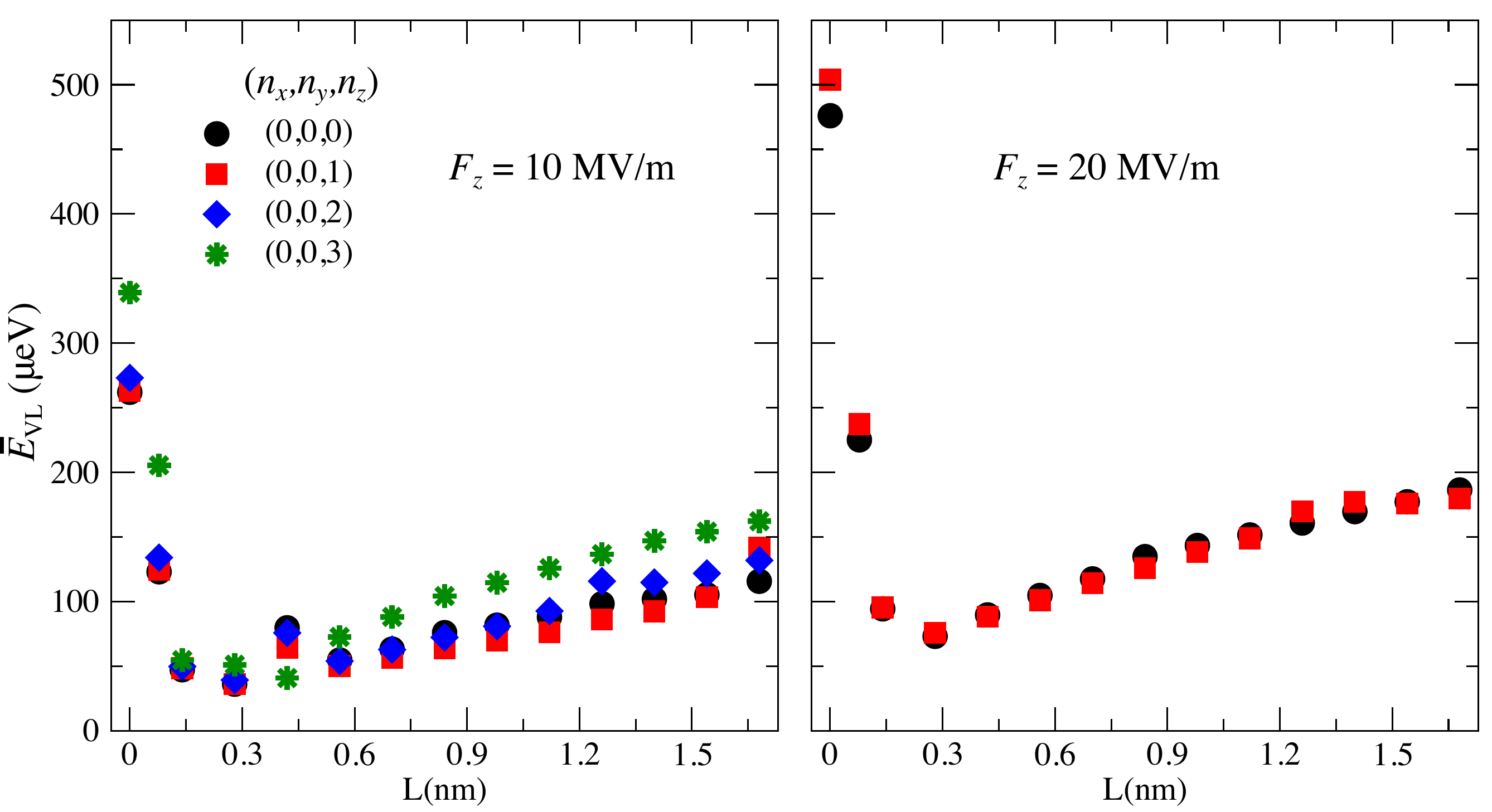}
\caption{\small The average valley splitting over $10^4$ realizations as a function of $L$ for the ground state and the excited states in the $\hat{z}$ direction that are confined at the quantum well. We consider two different electric fields $F_z$. In all cases, we see a strong initial suppression of the valley splitting, but after a specific width, the average valley splitting begins to increase again.} \label{vs}
\end{figure}

In Fig.~\ref{vs} we consider the average valley splitting as a function of the width of the interface for two values of electric field. We consider the case of the ground state being the QD qubit state and also the excited states in the $\hat{z}$ direction. We emphasize here that, e.g., the state $(0,0,1)$ is not the first excited state of the QD, since there are many other excited states related to the solutions in the $\hat{x}$ and $\hat{y}$ directions. Also, we calculated the VS here only for states confined to the Si quantum well. For this reason, we considered only states with energies below the energy of the state (0,0,2) for $F_z = 20$~MV/m.

The first value of $L$ different form zero that we considered is $L = 0.078$~nm, which means an interface of $0.4$~nm or 3 monolayers (ML). It is a value slightly smaller than the sharpest interface width realized experimentally so far (5 ML \cite{PaqueletWuetz2022}), but we believe that such sharpness can be reached in the foreseeable future. We can see that in all cases, there is a strong suppression of the average VS when we go from a perfect step interface ($L = 0$) to a smooth interface. For instance, a VS suppression greater than $60\%$ is obtained for the ground state case between $L = 0$ and $L = 0.078$~nm. This happens because the phase shift between the fast oscillations of the two low-lying valley states in silicon is averaged for a smooth interface, reducing the energy difference between them. 

However, we can see that the average VS begins to increase after a specific value of L.  This increase of the VS was mentioned recently in Ref.~\cite{PaqueletWuetz2022}, but a complete quantitative description of this behaviour was not reported before. This enhancement of the VS is due to the increase of the Ge content inside the Si well. 

\begin{figure}[]
\center
\includegraphics[width=0.9\linewidth]{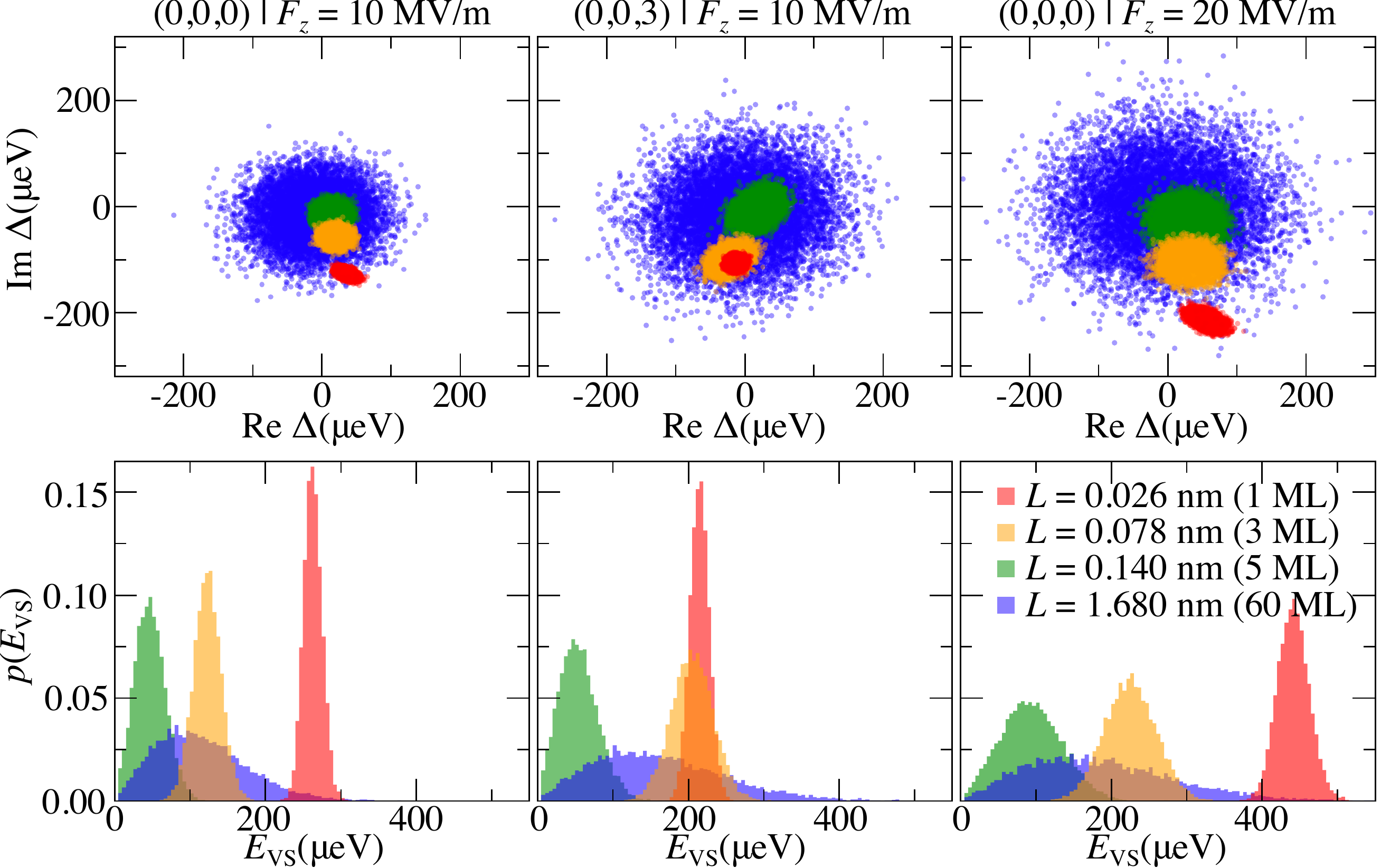}
\caption{\small Top panels: the distribution of the intervalley coupling $\Delta$ in the complex plane for $10^4$ realisations. Bottom panels: histogram of the probability density of the valley splitting. We consider here three cases: the ground state with $F_z = 10$~MV/m (left panels) and $F_z=20$~MV/m (right panels) and the state (0,0,3) with $F_z = 10$~MV/m (center panels). Each color represents a different interface width (see legend on the bottom right).} \label{RIp}
\end{figure}

We analyse the statistics of the VS in Fig.~\ref{RIp} for three different cases: the ground state with $F_z = 10$ and $20$~MV/m and the excited state (0,0,3) with $F_z = 10$~MV/m, where we plot the real and imaginary parts of the intervalley coupling $\Delta$ in the top panels and the probability density of the VS, $p(E_{VS})$, in the bottom panels. We consider four values of $L$. For the sake of comparison, one of this values is $L = 0.026$~nm, which is a unrealistic interface width of only one ML.   Note that QD excited states could be used for qubit realizations if the lower energy levels were completely filled.

In all cases, we have a Rician-like distribution for the real and imaginary parts of the intervalley coupling $\Delta$, where the distribution of points in the complex plane are circularly-symmetric. For a very sharp interface ($L = 0.026$~nm), the VS changes only for a small amount in each sample, which means a relatively deterministic distribution of VS, and we have a high VS value in each realisation. This is the desired scenario for silicon spin qubit, but such a narrow interface  is quite unrealistic. For a sharp but more realistic interface ($L = 0.078$~nm), we still have a fairly deterministic distribution of VS, but the values are reduced. Comparing the two ground state cases, we see that the electric field can shift up the distribution of VS. A similar effect is also achieved by considering the excited state (0,0,3) in comparison with the ground state. For $L = 0.14$~nm, the VS can be very small and the electric field cannot shift the distribution anymore, but only increase the VS average value by making the distribution wider. If we keep increasing the width of the interface, the distribution of VS becomes disorder-dominated, as can be seen for $L = 1.68$~nm, but with greater average VS. Therefore, the increase of the average VS obtained in Fig.~\ref{vs} is a consequence of a wider distribution of the VS. Even though the average VS for the cases $L = 0.078$~nm and $L=1.68$~nm are similar, the distribution of the VS values are very different. 

\begin{figure}[]
\center
\includegraphics[width=0.6\linewidth]{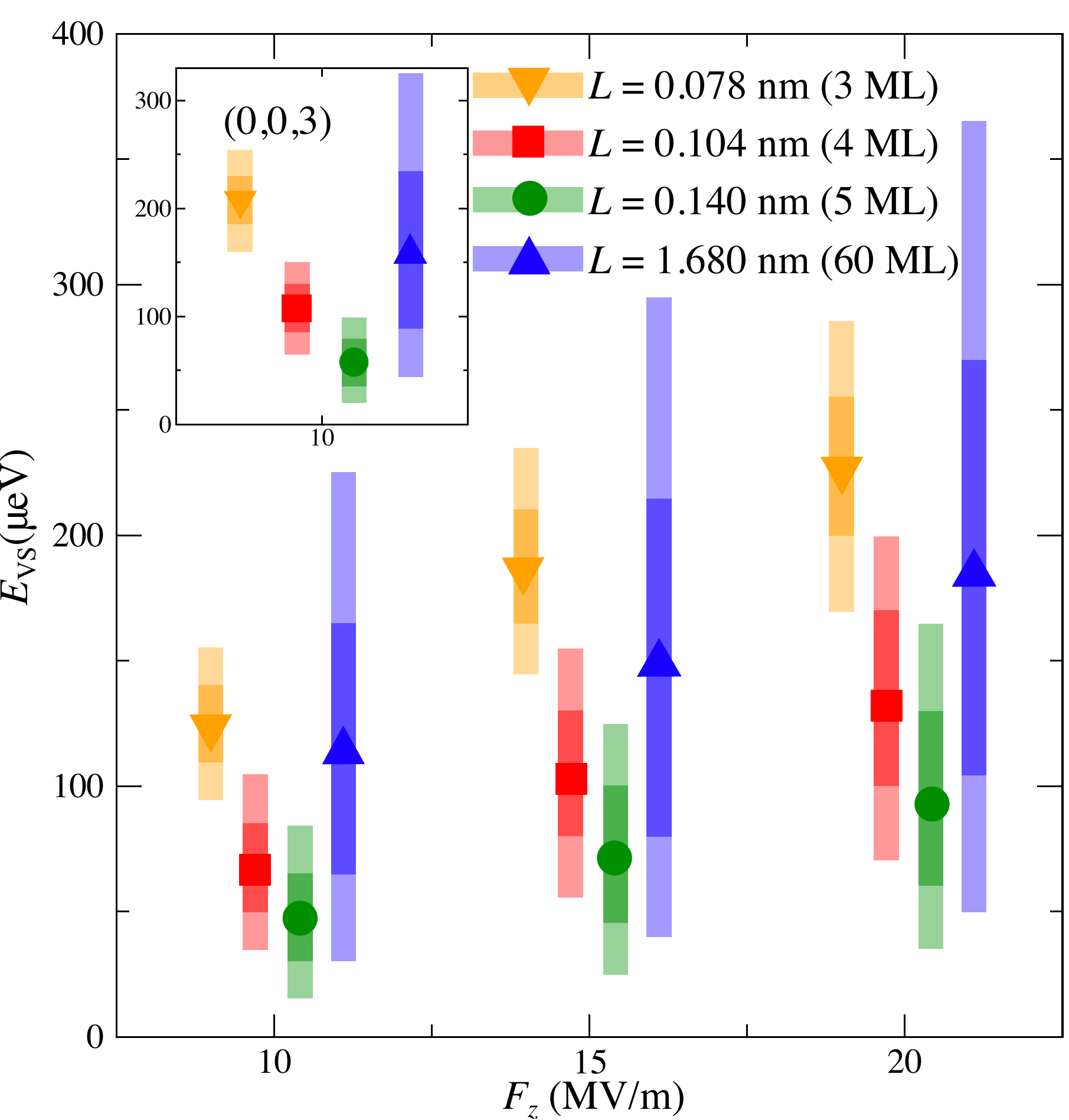}
\caption{\small Distribution of valley splitting for the ground state as function of the electric field for different interface width. The markers describe the average valley splitting, the lighter bars represent the 5-95 percentile range and the darker bars represent the 20-80 percentile range. Inset: distribution of valley splitting for the excited state (0,0,3) and $F_z = 10$~MV/m.} \label{porcentile}
\end{figure}

In principle, we are interested in more deterministic distributions of VS, since this implies that one has more control over its possible values. However, if the VS is distributed from very small values, the disorder-dominated cases are more suitable for spin qubit applications. This is confirmed when we look at 
Fig.~\ref{porcentile}, where we consider the distribution of VS as a function of the electric field for four values of $L$. The first three values of $L$ represent an interface width of $3$, $4$ and $5$~ML, respectively. The markers describe the average VS, the darker bars represent the 20-80 percentile range and the lighter bars represent the 5-95 percentile range. For $F_z=10$~MV/m we have a great percentile of the realisations with a VS below 100 $\mu$eV for almost all widths considered here, except for $L = 0.078$~nm. However, this is improved when we increase the electric field and also when we consider the excited state (0,0,3). For instance, for the excited state, our results predict that more than 95$\%$ of the devices should have a VS $\geq$ 160 $\mu$eV for $L=0.078$~nm, while more than 80$\%$ of the realisations will achieve a VS $\geq$ 85 $\mu$eV for $L=0.104$~nm and $L=1.68$~nm. When we increase the electric field to $20$~MV/m the results are even better, with more than 80$\%$ of the realisations achieving a VS $\geq$ 100 $\mu$eV for $L=0.104$~nm and $L=1.68$~nm and all devices having a VS $>$ 100 $\mu$eV for $L=0.078$~nm. The worst results were obtained for $L=0.14$~nm (interface width of $\approx 5$~ML). So, we can identify a width of 4 ML ($\sim 0.55$ nm) as a critical interface width, which means that for spin qubit application, the best option is an interface width no greater than 4 ML. However, if such sharp interfaces cannot be realised, it is better to consider the interface as wide as possible. 

\subsection{Valley splitting in the presence of a magnetic field}

\begin{figure}[]
\center
\includegraphics[width=0.9\linewidth]{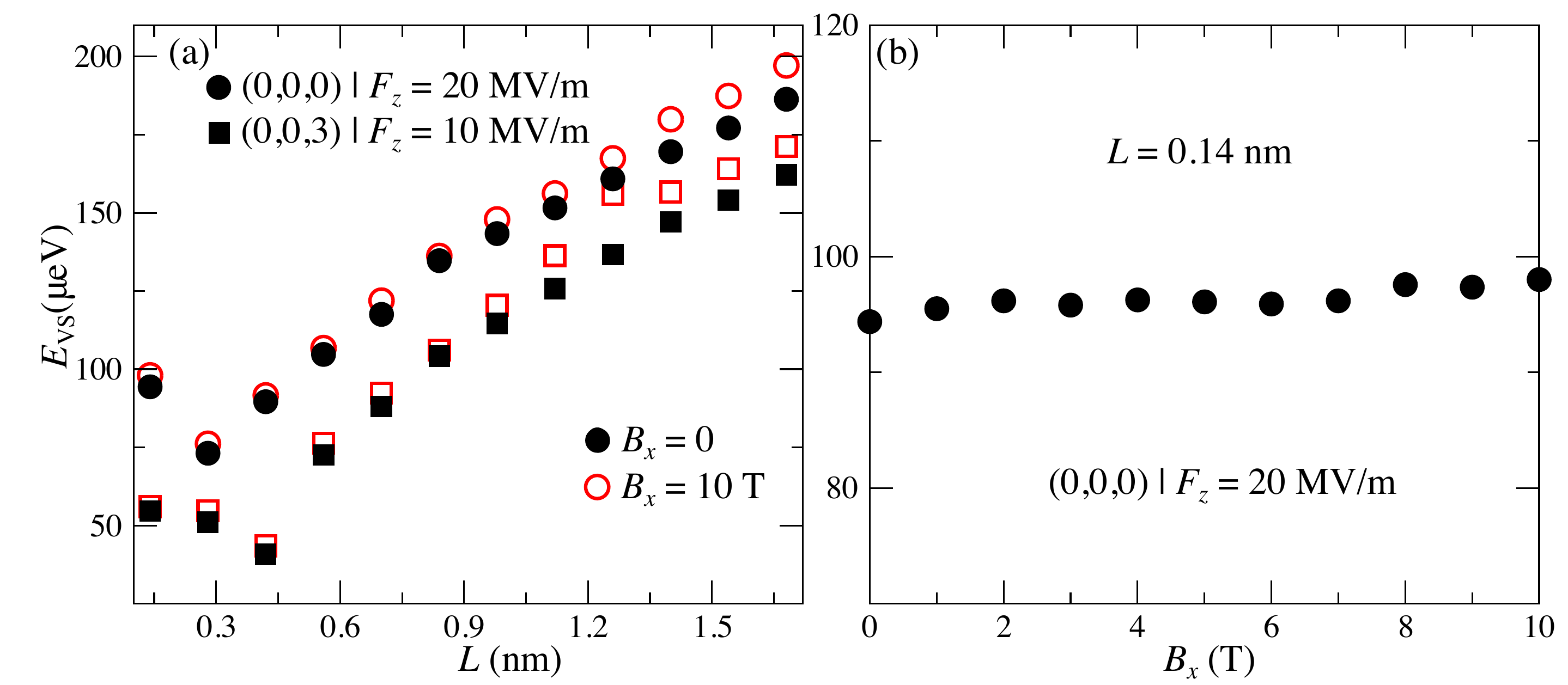}
\caption{\small (a) Average valley splitting as a function of $L$ for the ground state with $F_z = 20$~MV/m and the excited state (0,0,3) with $F_z = 10$ MV/m. We compare here the valley splitting with and without a magnetic field. (b) The average valley splitting as a function of a magnetic field along the $\hat{x}$ direction. We can see that the magnetic field has a very weak influence on the valley splitting.} \label{B}
\end{figure}

In Fig.~\ref{B} we can see the influence of the magnetic field on the valley splitting. We consider here that the magnetic field is in the $\hat{x}$ direction, which means that $B_y = 0$. In the left panel, we plot the average VS as a function of $L$ for two distinct qubit states with $B_x = 0$ and $B_x = 10$~T. We can see that such a strong magnetic field can only induce a small increase in the average VS, no matter if we consider the ground state or an excited state. 

This is confirmed in the right panel, where we consider the average VS as a function of the magnetic field for two interface widths. For $L=0.14$~nm, the average VS increases only 4$\%$ when the magnetic field goes from 0 to 10 T. For $L=1.54$~nm this increase is of 6.5$\%$. So, we can conclude that the magnetic field is not a good parameter for valley splitting engineering in Si-based heterostructures. 

\section{Conclusion}
\label{sec:conclusion}

To conclude, the lifetime of silicon spin qubits depends directly on the value of the valley splitting and the statistics of the valley splitting is very important for scalability. In this work, we develop a three dimensional model within the effective mass theory for the calculation of the valley splitting of realistic silicon-based heterostructures, which takes into account the alloy disorder at the interfaces and the lateral confinement. This model can be used to predict the valley splitting in silicon as a function of various parameters. Here, we consider the influence of the interface width and electromagnetic fields. Our results reveal that the electric field plays an important role in the distribution of valley splittings. For instance, for an interface width of 4 ML, we predict that 95$\%$ of the devices have a valley splitting $<$ 105 $\mu$eV for an electric field of $10$~MV/m, while 80$\%$ of the samples have a valley splitting $>$ 100 $\mu$eV when we consider an electric field of $20$~MV/m. For the width of the interface, we obtain that the best scenario for spin qubits is to fabricate devices with an interface $\leq$ 4 ML ($\sim 0.55$ nm). If such sharpness cannot be achieved, then the best option is to consider an interface as wide as possible. We also calculate the influence of the magnetic field, revealing that the magnetic field has a very weak effect on the valley splitting. Therefore, we can conclude that the magnetic field is not a good parameter for valley splitting engineering. We believe that our findings will contribute directly to a better fabrication of silicon-based heterostructures for spin qubits. A recent experimental realisation together with Nemo 3D tight-binding simulations reveal the change of the valley splitting as a function of the shape and location of the quantum dot \cite{McJunkin2022}. We plan to use our theoretical model to calculate this effect as well, since it is a calculation that cannot be done by any other model within the effective mass theory proposed before. 

\ack

This work has been funded by the Federal Ministry of Education and Research (Germany), funding code Grant No.~13N15657 (QUASAR). 

\appendix

\section{Potential at the Si/SiGe interface}
\label{sec:potential}

We are assuming here that the potential at the Si/SiGe interfaces is given by a distribution of delta functions in Eq.~(\ref{Ui}). Here, we will describe the 3D location of each delta function, the number of delta functions for each interface width and the calculation of the parameter $\lambda$.

\subsection{Location of the Ge atoms}

Since we are considering an elliptical quantum dot, the location of Ge atoms (delta functions) in the $xy$ plane for a uniform distribution is given by
\begin{eqnarray}
    x = 2x_0\sqrt{r}\cos{\theta}, \\
    y = 2y_0\sqrt{r}\sin{\theta},
\end{eqnarray}
where $r$ is a random value in the interval $[0,1]$ and $\theta$ is a random angle in the interval $[0,2\pi]$. The square root of $r$ ensures a uniform distribution in the ellipse, otherwise most of the Ge atoms would be in the middle of the ellipse. Also, we are taking into account only Ge atoms inside of an ellipse with twice of the semi-axis of the quantum dot. The contribution of extra Ge atoms to the valley splitting is negligibly small. 

In order to describe the distribution of Ge atoms at $\hat{z}$ direction, we take, e.g., the top interface, which is centered at $z = 0$. The distribution at the bottom interface can be obtained analogously. To model the smooth interface at $\hat{z}$ direction with a hyperbolic tangent function, we consider that the Ge atoms are located randomly in this direction following a probability distribution function (PDF) given by
\begin{eqnarray}
    \text{PDF} = \frac{1}{2}[\tanh(z/L_t)+1].
    \label{pdf}
\end{eqnarray}
We define the top interface region in the interval $-2.7 \cdot L < z < 2.7\cdot L$, which is an interval where the function in Eq.~(\ref{pdf}) is approximately in the range between 0 and 1.

\subsection{Number of Ge atoms}

With the volume of the interface region and taking into account that SiGe has approximately $5 \times 10^{22}$~atoms/cm$^3$, we can predict the number of delta functions, $N_{\rm Ge}$, in the interface potential (\ref{Ui}). In our case, this is done by considering that the Ge concentration changes at the interface from 0 to 30$\%$ following the hyperbolic tangent function in Eq.~(\ref{pdf}). The number of Ge atoms at the interface for some interface widths can be seen in Table~\ref{tab:Ge}.

\begin{table}
\begin{center}
%\captionof{table}{Total number of germanium atoms in the interface volume for some interface widths $L$.}
\caption{Total number of germanium atoms in the interface volume for some interface widths $L$.
\label{tab:Ge}}
    \begin{tabular}{|c|c|c|c|c|c|c|c|c|c|}
    \hline
       L(nm)  & 0.14 & 0.28 & 0.42 & 0.56 & 0.70 & 0.84 & 0.98 & 1.12 & 1.26 \\ \hline
       $N_{Ge}$ & 12825 & 25650 & 38475 & 51299 & 64124 & 76949 & 89774 & 102599 & 115424 \\
        \hline
    \end{tabular}
\end{center}
\end{table}

\subsection{The parameter $\lambda$}

We already know the number of Ge atoms at the interface and how we are modeling their 3D location. Let us now calculate the parameter $\lambda$. This is done by comparing the average potential (\ref{ubar}) with an average over a specific number of realization for the potential (\ref{Ui}). 

The integral of $\bar{U} = (U_0/2)[\tanh(z/L_t)+1]$ over the volume of the top interface gives
\begin{eqnarray}
    \frac{U_0}{2}V_I,
\end{eqnarray}
where $V_I$ is the volume of the interface. If we do the same for the average over a specific number of realizations for the potential (\ref{Ui}), this integral gives $\lambda N_{Ge}$. This means that
\begin{eqnarray}
    \lambda = \frac{U_0 V_I}{2N_{Ge}} = 10 \; \text{meV}\cdot \text{nm}.
\end{eqnarray}

\begin{figure}[]
\center
\includegraphics[width=0.7\linewidth]{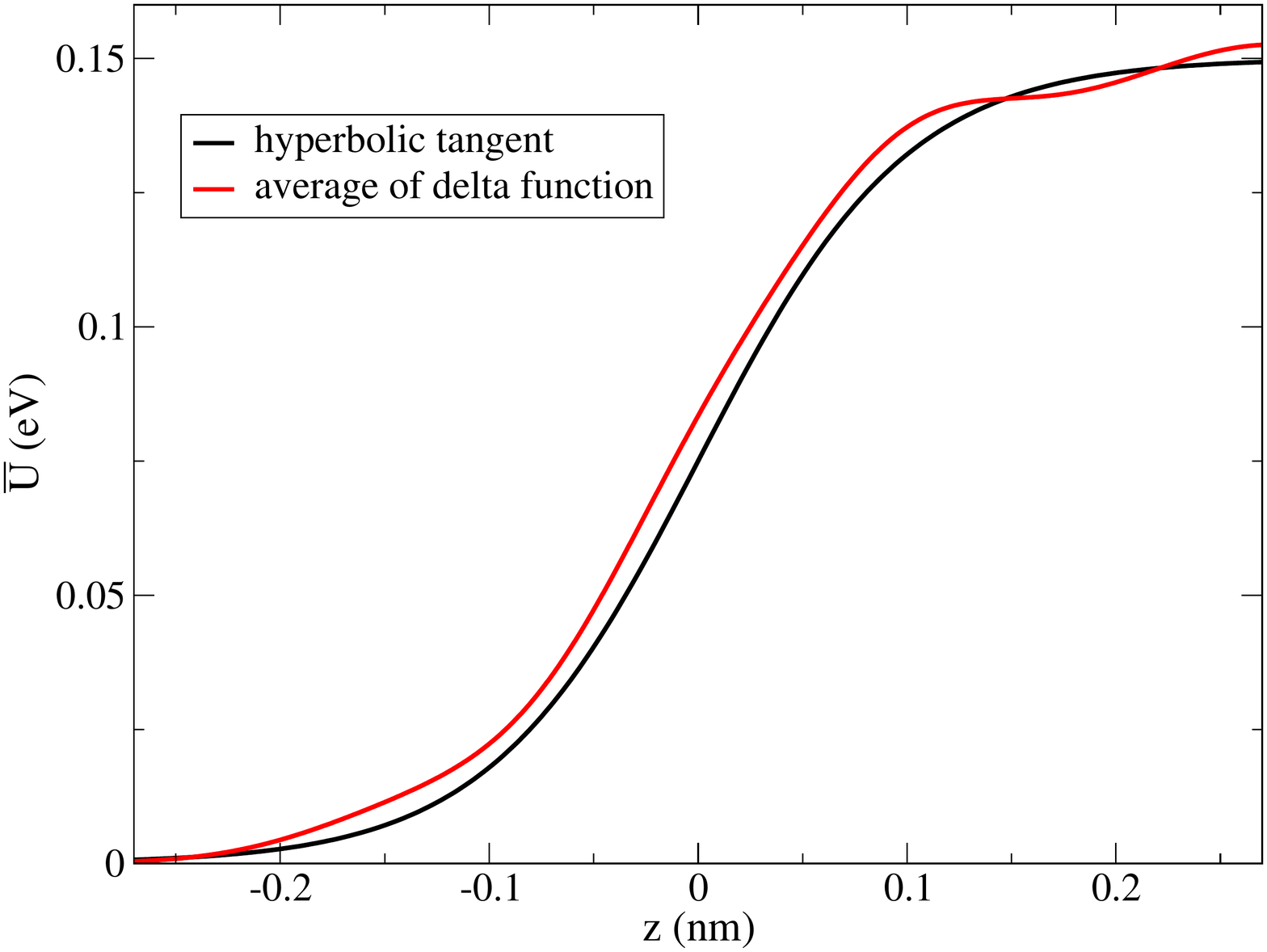}
\caption{\small Comparison between the average potential (\ref{ubar}) (black line) with the potential (\ref{Ui}) (red line) obtained over 10$^4$ realizations for $L = 0.1$~nm.} \label{U}
\end{figure}

In Fig.~\ref{U} we compare the average potential (\ref{ubar}) with the potential (\ref{Ui}) obtained over 10$^4$ realizations for $L = 0.1$~nm. The plot was possible by considering a smeared-out delta function, where we replace each delta function in the following way
\begin{eqnarray}
    \delta(x-x_0) \rightarrow \lim_{\epsilon \rightarrow 0} \frac{1}{2\sqrt{\pi \epsilon}}e^{-(x-x_0)^2/(4\epsilon)}.
\end{eqnarray}
We numerially used $\epsilon = 10^{-21}$ for our calculations.

\section{Numerical calculation of eigenenergies and envelope functions}
\label{sec:numeric}

The Eq.~(\ref{sch2}) has an analytical solution only when $\tilde{U}$ is constant. This is the case within the lower and upper SiGe barrier regions, as well as at the Si well. We solve this equation using the transfer matrix method, which means that for the top and bottom interfaces, where $\tilde{U}$ depends on position, we decompose the potential in successive constant rectangular barriers. In our calculations, we decompose each Si/SiGe interface into 500 small regions. In each region of the heterostructure, including the small regions at the top and bottom interfaces, the solution is given by
\begin{eqnarray}
    \psi_{z,n_z}^j = \frac{N}{\sqrt{z_0}}\left[c_j \text{Ai}(\tilde{U}_j-\tilde{z}-\tilde{\epsilon}_{z,n_z}) + d_j \text{Bi}(\tilde{U}_j-\tilde{z}-\tilde{\epsilon}_{z,n_z})\right],
\end{eqnarray}
where Ai and Bi are the Airy functions of first and second kinds, respectively, $c_j$ and $d_j$ are constants to be determined, $N$ is a normalisation constant and $j$ labels all the regions. 

We can write the continuity of the envelope function and its first derivative at a specific interface between two regions, say regions 1 and 2, as follows
\begin{eqnarray}
   M_1(\tilde{z}_{12})\left(\begin{array}{cc}c_1 \\ d_1 \end{array}\right)
    =M_2(\tilde{z}_{12}) \left(\begin{array}{cc}c_2 \\ d_2 \end{array}\right),
\end{eqnarray}
where $\tilde{z}_{12} = z_{12}/z_0$, $z_{12}$ is the location in the $\hat{z}$ axis of the interface between regions 1 and 2 and
\begin{eqnarray}
    M_j(\tilde{z}) =  \left(\begin{array}{cc}\text{Ai}(\tilde{U}_j-\tilde{z}-\tilde{\epsilon}_{z,n_z}) & \text{Bi}(\tilde{U}_j-\tilde{z}-\tilde{\epsilon}_{z,n_z}) \\ \text{Ai}^{\prime}(\tilde{U}_j-\tilde{z}-\tilde{\epsilon}_{z,n_z}) & \text{Bi}^{\prime}(\tilde{U}_j-\tilde{z}-\tilde{\epsilon}_{z,n_z}) \end{array}\right).
\end{eqnarray}
Hence we find that
\begin{eqnarray}
    \left(\begin{array}{cc}c_1 \\ d_1 \end{array}\right)
    = M_{12} \left(\begin{array}{cc}c_2 \\ d_2 \end{array}\right),
\end{eqnarray}
where $M_{12} = M_1^{-1}(\tilde{z}_{12})M_2(\tilde{z}_{12})$. If we label the lower and upper SiGe barriers as regions $i$ and $f$, respectively, we can write
\begin{eqnarray}
    \left(\begin{array}{cc}c_i \\ 0 \end{array}\right)
    = M \left(\begin{array}{cc}c_f \\ d_f \end{array}\right),
    \label{m}
\end{eqnarray}
where 
\begin{eqnarray}
    M = M_{i1}M_{12}M_{23}\text{...}M_{nf}
\end{eqnarray}
and we consider $d_i = 0$, since the Bi function does not decay inside of the lower barrier. 

From Eq.~(\ref{m}), we have that
\begin{eqnarray}
    m_{21}c_f + m_{22}d_f = 0,
\end{eqnarray}
where $m_{ij}$ are the coefficients for the matrix $M$. We also have that at $z = d_b$ the envelope function has to be equal to zero. Therefore,
\begin{eqnarray}
    c_f \text{Ai}(\zeta_b) + d_f \text{Bi}(\zeta_b) = 0,
\end{eqnarray}
where $\zeta_b = U_0/\epsilon_0 - d_b - \tilde{\epsilon}_{z,n_z}$. From the last two equations, we conclude that
\begin{eqnarray}
    m_{22}-m_{21}\frac{\text{Bi}(\zeta_b)}{\text{Ai}(\zeta_b)}= 0.
    \label{en}
\end{eqnarray}

The eigenenergies are obtained from Eq.~(\ref{en}). Once we have these energies, we can obtain all the coefficients $c_j$ and $d_j$ by solving a system of equations. Finally, with the energies and the coefficients, we can normalise the envelope function and obtain the constant $N$.

\section*{References}

\bibliographystyle{iopart-num}
\bibliography{ref.bib}

\end{document}